\documentclass[aps,pre,floats,twocolumn,showpacs,superscriptaddress]{revtex4}

\usepackage{natbib}
\usepackage[utf8]{inputenc}

\usepackage{amsmath, amsthm, amssymb}
\usepackage{enumitem}
\usepackage{graphics,graphicx}
\usepackage{epsfig}
\usepackage{times}
\usepackage{dcolumn}
\usepackage{url}
\usepackage{color}

\usepackage[all]{xy}
\usepackage{mdwlist}

\begin{document}

\renewcommand*\thesection{\arabic{section}}
\newcommand{\beq}{\begin{equation}}
\newcommand{\eeq}{\end{equation}}
\newcommand{\sss}{\scriptscriptstyle}
\newcommand{\sz}{\scriptsize}
\newcommand{\rev}[1]{{\color{black} #1}}
\newcommand{\rerev}[1]{{\color{red} #1}}
\newcommand{\bea}{\begin{eqnarray}}
\newcommand{\eea}{\end{eqnarray}}
\newcommand{\nn}{\nonumber \\}

\newcommand{\pus}[1]{{p}_{#1}^{\mbox{\sz US}}}
\newcommand{\pas}[1]{{p}_{#1}^{\mbox{\sz AS}}}
\newcommand{\pai}[1]{{p}_{#1}^{\mbox{\sz AI}}}
\newcommand{\pui}[1]{{p}_{#1}^{\mbox{\sz UI}}}
\newcommand{\qu}[1]{{q}_{#1}^{\mbox{\sz U}}}
\newcommand{\qa}[1]{{q}_{#1}^{\mbox{\sz A}}}
\newcommand{\paw}[1]{{p}_{#1}^{\mbox{\sz A}}}
\newcommand{\pinf}[1]{{p}_{#1}^{\mbox{\sz I}}}
\newcommand{\psus}[1]{{p}_{#1}^{\mbox{\sz S}}}
\newcommand{\pu}[1]{{p}_{#1}^{\mbox{\sz U}}}
\newcommand{\pa}[1]{{p}_{#1}^{\mbox{\sz A}}}
\newcommand{\rhoinf}{{\rho}^{\mbox{\sz I}}}
\newcommand{\rhoui}{{\rho}^{\mbox{\sz UI}}}
\newcommand{\rhoai}{{\rho}^{\mbox{\sz AI}}}
\newcommand{\rhoaw}{{\rho}^{\mbox{\sz A}}}
\newcommand{\betau}{{\beta}^{\mbox{\sz U}}}
\newcommand{\betaa}{{\beta}^{\mbox{\sz A}}}
\newcommand{\pust}[1]{{p}_{#1}^{\mbox{\sz US}}(t)}
\newcommand{\past}[1]{{p}_{#1}^{\mbox{\sz AS}}(t)}
\newcommand{\pait}[1]{{p}_{#1}^{\mbox{\sz AI}}(t)}
\newcommand{\puit}[1]{{p}_{#1}^{\mbox{\sz UI}}(t)}
\newcommand{\qt}[1]{{q}_{#1}(t)}
\newcommand{\qut}[1]{{q}_{#1}^{\mbox{\sz U}}(t)}
\newcommand{\qat}[1]{{q}_{#1}^{\mbox{\sz A}}(t)}
\newcommand{\pawt}[1]{{p}_{#1}^{\mbox{\sz A}}(t)}
\newcommand{\pinft}[1]{{p}_{#1}^{\mbox{\sz I}}(t)}
\newcommand{\psust}[1]{{p}_{#1}^{\mbox{\sz S}}(t)}
\newcommand{\pustt}[1]{{p}_{#1}^{\mbox{\sz US}}(t+1)}
\newcommand{\pastt}[1]{{p}_{#1}^{\mbox{\sz AS}}(t+1)}
\newcommand{\paitt}[1]{{p}_{#1}^{\mbox{\sz AI}}(t+1)}
\newcommand{\puitt}[1]{{p}_{#1}^{\mbox{\sz UI}}(t+1)}
\newcommand{\pinftt}[1]{{p}_{#1}^{\mbox{\sz I}}(t+1)}
\newcommand{\psustt}[1]{{p}_{#1}^{\mbox{\sz S}}(t+1)}

\definecolor{MyColorito}{rgb}{1,0.5,0.5}
\definecolor{MyColorito2}{rgb}{0.1,0.2,1}
\definecolor{MyColorito3}{rgb}{0.1,0.1,0.1}
\newcommand{\ClaraMetaComment}[1]{\textcolor{MyColorito}{[\textbf{CLARA:} #1]}}
\newcommand{\clara}[1]{\textcolor{MyColorito}{#1}}
\newcommand{\blau}[1]{\textcolor{MyColorito3}{#1}}
\newcommand{\hola}[1]{\textcolor{MyColorito2}{#1}}

\title{Competing spreading processes on multiplex networks: awareness and epidemics}

\author{Clara Granell}
\affiliation{Departament d'Enginyeria Inform\`atica i Matem\`atiques, Universitat Rovira i Virgili, 43007 Tarragona, Spain}

\author{Sergio G\'omez}
\affiliation{Departament d'Enginyeria Inform\`atica i Matem\`atiques, Universitat Rovira i Virgili, 43007 Tarragona, Spain}

\author{Alex Arenas}
\affiliation{Departament d'Enginyeria Inform\`atica i Matem\`atiques, Universitat Rovira i Virgili, 43007 Tarragona, Spain}

\begin{abstract}
Epidemic-like spreading processes on top of multilayered interconnected complex networks reveal a rich phase diagram of intertwined  competition effects.
A recent study by the authors [Granell et al. Phys. Rev. Lett. 111, 128701 (2013)] presented the analysis of the interrelation between two processes accounting for the spreading of an epidemics, and the spreading of information awareness to prevent its infection, on top of multiplex networks. The results in the case in which awareness implies total immunization to the disease, revealed the existence of a metacritical point at which the critical onset of the epidemics starts depending on the reaching of the awareness process. Here we present a full analysis of these critical properties in the more general scenario where the awareness spreading does not imply total immunization, and where infection does not imply immediate awareness of it. We find the critical relation between both competing processes for a wide spectrum of parameters representing the interaction between them. We also analyze the consequences of a massive broadcast of awareness (mass media) on the final outcome of the epidemic incidence. Importantly enough, the mass media makes the metacritical point to disappear. The results reveal that the main finding i.e.\ existence of a metacritical point, is rooted on the competition principle and holds for a large set of scenarios.
\end{abstract}


\maketitle

\section{Introduction}

During the last years significant progress has been made in the study of complex networks, to the point that we now have a comprehensive toolset to characterize them \cite{Newman2010}. This recent flurry of activity in network science has been induced by increased computing power and by the possibility of studying the properties of large databases describing real networks. The challenge remains, however, making network theory predictive, that is, developing the methods that will help us turn network data into quantitative predictions for complex systems. Importantly, these methods need to take into account that, as it is increasingly recognized, complex systems are often composed of interacting layers, giving rise to multilevel networks (see \cite{2013arXiv1309.7233K} and references therein). For example, cellular processes are the result of the interaction of metabolic networks, protein interaction networks, and gene regulation networks, among others; social networks are formed by individuals who are interconnected at many different levels and whose connectivity varies with time; and transportation networks are the superposition of different transportation services.

A particularly interesting setup comes from those multilayer interconnected networks in which the nodes represent the same entities in all layers, these networks have been usually called {\em multiplex} networks \cite{kurant06,mucha10,Szell10,gardenes12,baxter12,cozzo12,bianconi13,gomez13,soleribalta13,radicchi13,dedomenico13,SahnehSM13,sanz2014}. The understanding of the emergent physical phenomena on multiplex networks is gaining much attention as a particular case of interdependent networks \cite{buldyrev10,gao11}. An archetypical example of a multiplex network is a social network in which the different layers represent different types of social relationships. For example, one can place friendship ties, family ties, and coworker ties in three different layers. Note that this scenario is particularly interesting for epidemics spreading, raising the question about what is the outcome of the epidemics given that we have several layers for its spreading, and more intriguing, the interplay between awareness and epidemics when both phenomena compete using different layers of propagation. By understanding this interplay one can assess the consequences the awareness can have on the outbreak of the epidemics and its incidence \cite{funk09}.

Recently, the authors investigated the interplay between awareness and epidemic spreading in multiplex networks \cite{granell13} in the particular scenario where infection of the epidemics implies immediate awareness and awareness implies total immunization of the epidemics. Here we relax these two strong assumptions and investigate the consequences. The two parameters, self-awareness  and degree of immunization, are regulated by probabilities $\kappa$ and $\gamma$ respectively. We find that while the self-awareness does not affect the critical properties of the system, the degree of immunization does.
In this generalized model we also include the effect of massive awareness information flowing through the network, the {\em mass media effect}. We assess whether an external node which represents the mass media (TV, radio, newspapers, etc.), connected to all nodes in the information layer, regularly transmitting information about the disease is crucial to the final outcome of the epidemics. The findings show that the presence of the mass media makes the metacritical point of the epidemics vanish.

The paper is organized as follows: in the next section we expose the model presented in its extended form, including the mass media. In Section 3, we present the Microscopic Markov Chain Approach to analytically represent the previous model. In Section 4 we find analytically the onset of the epidemics, and in Section 5 we present the results. Section 6 is devoted to discussion and conclusions.

\section{Model for awareness and epidemic spreading with mass media}

Dynamical processes in real world are not isolated but they interact each other. Epidemic spreading, for instance, can be effectively represented as a diffusive process in a single layer, using well-known models such as Susceptible-Infected-Susceptible (SIS), Susceptible-Infected-Recovered (SIR), and others. However, this particular spreading could be affected by other processes, such as diffusion of information about the epidemics itself \cite{Meloni11}.

For example, epidemics such as flu and other contagious diseases may spread in a network representing the physical contacts people have. People you see every day in the office, family and close friends are nodes of this network. On the other hand, consider a network of information, whose links now represent the relations with people you regularly share information with. Some of these entities will be the same as in the physical contacts network, and others are just sources of information such as friends in Twitter or any other online social network to whom you do not have a regular physical contact. When there is an important outbreak of an epidemics, for instance seasonal flu, or any other special event involving a disease, information about the presence of this illness is spread in the social networks. Awareness of a disease often leads to individuals taking preventive measures in the physical contact layer \cite{influenzanet}.

Here we describe a setup involving two competing spreading processes: the spreading of information holds back the spreading of the disease, while the nodes infected by the disease support the information spreading process by generating new aware individuals. The abstracted model is then as follows: consider a multiplex network formed by two layers; the bottom layer is formed by the network of physical contacts, while the top one is a representation of an online social network. All nodes represent the same entities in both layers, but the connectivity is different in each of them. On top of the physical contacts layer we assimilate a Susceptible-Infected-Susceptible (SIS) process, where the probability that a susceptible node gets infected after a contact with another infected node is $\beta$, and the probability of an infected node to spontaneously recover is $\mu$. On the awareness layer we apply an equivalent process Unaware-Aware-Unaware (UAU), with parameters $\lambda$ and $\delta$ playing the roles of $\beta$ and $\mu$, respectively.

The interaction between both processes is modeled as follows: a node that is infected in the SIS layer will become Aware in the UAU layer with probability $\kappa$. This probability accounts for the possibility that the nodes may not know they are infected or may choose not to spread information about it. Similarly, a node that is aware on the UAU layer will take measures for preventing infection, therefore the parameter $\gamma$ regulates the probability of a node to get infected. The infectivity parameter of the SIS is then different depending on the state of the node in the information layer. $\betau$ regulates the probability of a node to get infected when it is unaware of the disease, while $\betaa=\gamma \betau$ regulates the probability when the node is aware. Thus, the parameter $\gamma$ ranges from 0, representing total immunization, to 1, representing no effect of the information awareness on the epidemics. Equivalently, we can regulate the upwards interaction by tuning the parameter $\kappa$ from 0 to 1. Note that when $\gamma=1$ and $\kappa=0$ the two interactions are disabled and the setup becomes equivalent to running both processes in single layer independent networks.

\begin{figure}[tb]
  \begin{center}
  \includegraphics[width=8cm,clip=]{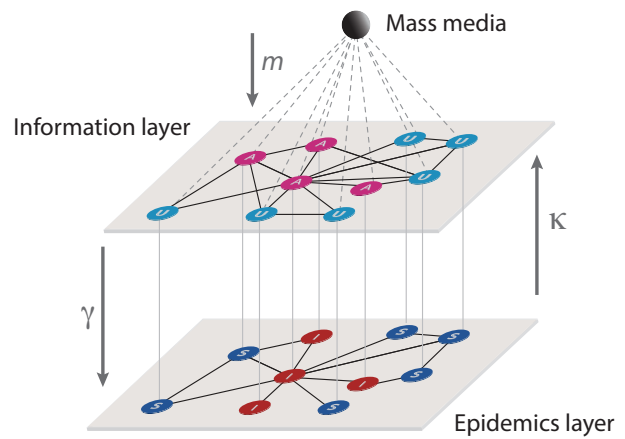}
  \end{center}
  \caption{(color online). Awareness-epidemic model in the presence of mass media. The upper (information) layer is supporting the spreading of awareness, and nodes have two possible states: unaware (U) or aware (A). The lower (epidemic) layer corresponds to the network where the epidemic spreading takes place. The nodes are the same actors than in the upper layer, but here their state can be: susceptible (S) or infected (I). The mass media is represented as a top node that provides with information to the full system.}
  \label{fig:sketch}
\end{figure}

The proposed model above is restricted to local flow of information, a word-of-mouth propagation in the information layer. Nevertheless, a certainly more realistic scenario is prescribed when considering that information can also have global impact on the system, this is the usual case when we consider the  effect of mass media. Mass media are entities with eventually a large impact (connectivity) that regularly transmit information all over the entire population. Moreover, they are perceived as reliable sources of information, and therefore its role when warning from an epidemics outbreak may be crucial to the final outcome of the disease.

To incorporate this effect to our setup, we add a single node connected to all nodes in the UAU layer, that will regularly transmit information of the presence of a disease and each individual gets aware with probability $m$. In Fig.~\ref{fig:sketch} we depict a sketch of the resulting scenario.

\section{Microscopic Markov Chain Approach}

Summing up, the $N$ nodes in the multiplex model proposed can be in the following states: US (Unaware and Susceptible), UI (Unaware and Infected), AS (Aware and Susceptible) or AI (Aware and Infected). A methodological way to discover the dynamical equations governing the system is to build first discrete transition probability trees that account for all the possible changes of state (and their probabilities) at every time step. Let us illustrate how to build these trees using a stand-alone SIS process, with nodes in states either Susceptible (S) or Infected (I). As explained above, $\beta$ represents the probability that a susceptible node becomes infected after a contact with one of its infected neighbors, and $\mu$ is the recovery probability for infected nodes. In a standard SIS model (reactive process \cite{Colizza:2007,gleeson11}), each infected node contacts all its neighbors at each time step, thus it is convenient to define the probability $1-q_i$ that a susceptible node $i$ gets infected by at least one of its infected neighbors. Conversely, $q_i$ represents the probability that none of the neighbors of $i$ infects it. The possible changes of state of the nodes and their probabilities at every time step can be represented by the following state transition trees:
\[
\begin{array}{lll}
  \xymatrixcolsep{5mm}
  \xymatrixrowsep{1mm}
  \vcenter{\xymatrix{
    && S
    \\
    I \ar[urr]^{\mbox{$\mu$}} \ar[drr]_{\mbox{$1-\mu$}}
    \\
    && I
  }}
  &
  \mbox{}\hspace{1cm}\mbox{}
  &
  \xymatrixcolsep{5mm}
  \xymatrixrowsep{1mm}
  \vcenter{\xymatrix{
    && S
    \\
    S \ar[urr]^{\mbox{$q_i$}} \ar[drr]_{\mbox{$1-q_i$}}
    \\
    && I
  }}
\end{array}
\]
The roots of the trees represent the possible states of a node at time $t$, hence the need of two trees, one for state I and another for state S. The leaves of each tree account for the possible states at time $t+1$. The transition arrows are labeled with the corresponding probabilities, and they may depend on the node (e.g.\ $q_i$) and also on the time step $t$; this time dependence has not been made explicit in the trees for the sake of simplicity.

From these transition trees it is possible to recover the Microscopic Markov Chain Approach (MMCA) equations \cite{gomez10} which express the probability of a node being in each state at time $t+1$ as a function of its state in the previous time step. For example, the probability $\pinftt{i}$ of node $i$ being infected (I) at time $t+1$ has two contributions, one for each branch in which state $I$ appears as a leaf in the trees. From the left tree we get the contribution $\pinft{i} (1-\mu)$, which corresponds to the case in which the node was infected (I) and has not recovered, and from the tree in the right we get $\psust{i} (1-\qt{i})$, which accounts for the case in which the node was healthy (S) but has been infected by any of its neighbors. After doing the same procedure for the branches ending in state S, the final MMCA equations read
\bea
  \pinftt{i} &=& \pinft{i} (1-\mu) + \psust{i} (1-\qt{i})\,, \\
  \psustt{i} &=& \pinft{i} \mu + \psust{i} \qt{i}\,.
\eea
These two equations fulfill, for all time steps, the normalization condition $\pinf{i}+\psus{i}=1$, thus only one of them is really needed, which is the standard MMCA equation for the SIS process:
\beq
  \pinftt{i} = \pinft{i} (1-\mu) + (1-\pinft{i}) (1-\qt{i})\,. \\
\eeq

\begin{figure*}[htbp]
  \begin{center}
  \begin{tabular}{lll}
    a) & \hspace{5mm} & b)
  \\

  $
  \xymatrixcolsep{5mm}
  \xymatrixrowsep{1mm}
  \vcenter{\xymatrix{
    && && && AS
    \\
    && && AI \ar[urr]^{\mbox{$\mu$}} \ar[drr]_{\mbox{$1-\mu$}}
    \\
    && && && AI
    \\
    && UI \ar[uurr]^{\mbox{$m$}} \ar[ddrr]_{\mbox{$1-m$}}
    \\
    && && && US
    \\
    && && UI \ar[urr]^{\mbox{$\mu$}} \ar[drr]_{\mbox{$1-\mu$}} && && AI
    \\
    UI \ar[uuurr]^{\mbox{$r_i$}} \ar[dddrr]_{\mbox{$1-r_i$}}
      && && && UI \ar[urr]^{\mbox{$\kappa$}} \ar[drr]_{\mbox{$1-\kappa$}}
    \\
    && && && && UI
    \\
    && && && AS
    \\
    && AI \ar[rr]^{\mbox{$1$}} && AI \ar[urr]^{\mbox{$\mu$}} \ar[drr]_{\mbox{$1-\mu$}}
    \\
    && && && AI
  }}
  $

  & &

  $
  \xymatrixcolsep{5mm}
  \xymatrixrowsep{1mm}
  \vcenter{\xymatrix{
    && && && AS
    \\
    && && AI \ar[urr]^{\mbox{$\mu$}} \ar[drr]_{\mbox{$1-\mu$}}
    \\
    && && && AI
    \\
    && UI \ar[uurr]^{\mbox{$m$}} \ar[ddrr]_{\mbox{$1-m$}}
    \\
    && && && US
    \\
    && && UI \ar[urr]^{\mbox{$\mu$}} \ar[drr]_{\mbox{$1-\mu$}} && && AI
    \\
    AI \ar[uuurr]^{\mbox{$\delta$}} \ar[dddrr]_{\mbox{$1-\delta$}}
      && && && UI \ar[urr]^{\mbox{$\kappa$}} \ar[drr]_{\mbox{$1-\kappa$}}
    \\
    && && && && UI
    \\
    && && && AS
    \\
    && AI \ar[rr]^{\mbox{$1$}} && AI \ar[urr]^{\mbox{$\mu$}} \ar[drr]_{\mbox{$1-\mu$}}
    \\
    && && && AI
  }}
  $

  \\ \\
    c) & & d)
  \\

  $
  \xymatrixcolsep{5mm}
  \xymatrixrowsep{1mm}
  \vcenter{\xymatrix{
    && && && AS
    \\
    && && AS \ar[urr]^{\mbox{$q_i^A$}} \ar[drr]_{\mbox{$1-q_i^A$}}
    \\
    && && && AI
    \\
    && US \ar[uurr]^{\mbox{$m$}} \ar[ddrr]_{\mbox{$1-m$}}
    \\
    && && && US
    \\
    && && US \ar[urr]^{\mbox{$q_i^U$}} \ar[drr]_{\mbox{$1-q_i^U$}} && && AI
    \\
    US \ar[uuurr]^{\mbox{$r_i$}} \ar[dddrr]_{\mbox{$1-r_i$}}
      && && && UI \ar[urr]^{\mbox{$\kappa$}} \ar[drr]_{\mbox{$1-\kappa$}}
    \\
    && && && && UI
    \\
    && && && AS
    \\
    && AS \ar[rr]^{\mbox{$1$}} && AS \ar[urr]^{\mbox{$q_i^A$}} \ar[drr]_{\mbox{$1-q_i^A$}}
    \\
    && && && AI
  }}
  $

  & &

  $
  \xymatrixcolsep{5mm}
  \xymatrixrowsep{1mm}
  \vcenter{\xymatrix{
    && && && AS
    \\
    && && AS \ar[urr]^{\mbox{$q_i^A$}} \ar[drr]_{\mbox{$1-q_i^A$}}
    \\
    && && && AI
    \\
    && US \ar[uurr]^{\mbox{$m$}} \ar[ddrr]_{\mbox{$1-m$}}
    \\
    && && && US
    \\
    && && US \ar[urr]^{\mbox{$q_i^U$}} \ar[drr]_{\mbox{$1-q_i^U$}} && && AI
    \\
    AS \ar[uuurr]^{\mbox{$\delta$}} \ar[dddrr]_{\mbox{$1-\delta$}}
      && && && UI \ar[urr]^{\mbox{$\kappa$}} \ar[drr]_{\mbox{$1-\kappa$}}
    \\
    && && && && UI
    \\
    && && && AS
    \\
    && AS \ar[rr]^{\mbox{$1$}} && AS \ar[urr]^{\mbox{$q_i^A$}} \ar[drr]_{\mbox{$1-q_i^A$}}
    \\
    && && && AI
  }}
  $
  \end{tabular}
  \end{center}
  \caption{Transition probability trees for each one of four possible states a node may be in. The root of each tree represents the state of any node at time $t$, and the leaves their states at time $t+1$. Each time step is subdivided in four phases: awareness spreading (UAU process), mass media broadcast, epidemic spreading (SIS process) and self-awareness of being infected.}
  \label{fig:arbres}
\end{figure*}
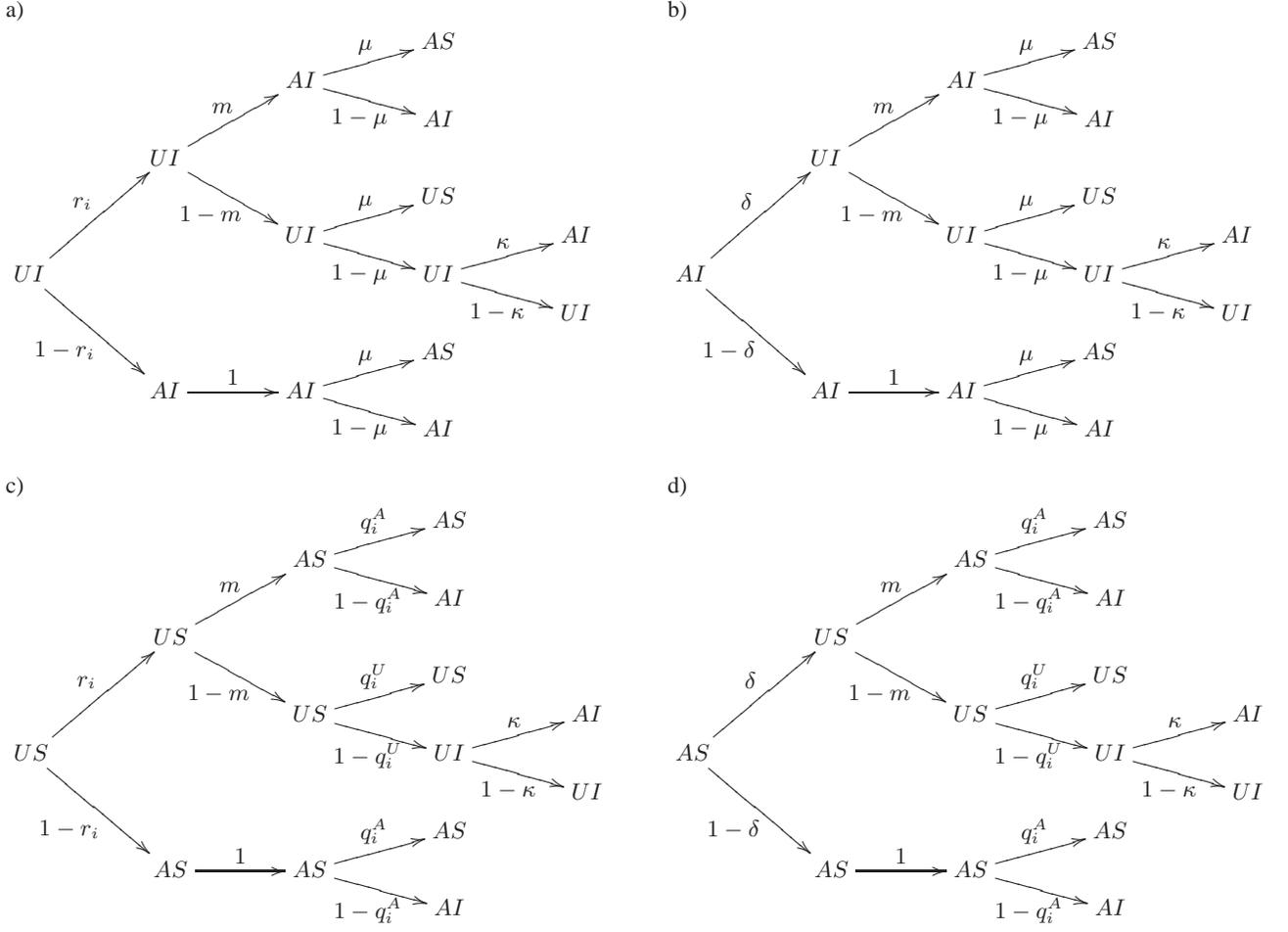

Following this procedure we can deal with more complex situations as the proposed model of competing awareness and epidemic spreading with mass media. The transition trees for our model are shown in Fig.~\ref{fig:arbres}. They are structured according to the four phases in which every time step is divided: awareness spreading (UAU process), mass media broadcast, epidemic spreading (SIS process) and self-awareness of being infected. The resulting MMCA equations representing the probabilities of every node node being in each of the four possible states are:
\begin{widetext}
\bea
  \pustt{i} &=& \puit{i} r_i(t) (1-m)\mu
                + \pait{i}\delta (1-m)\mu
                + \pust{i}r_i(t) (1-m)\qut{i}
                + \past{i} \delta (1-m)\qut{i}\,, \label{eq:MMCApus} \\
  \puitt{i} &=& \puit{i} r_i(t) (1-m)(1-\mu)(1-\kappa)
                + \pait{i}\delta (1-m)(1-\mu)(1-\kappa) \label{eq:MMCApui} \\
                && \mbox{} + \pust{i} r_i(t) (1-m)(1-\qut{i})(1-\kappa)
                + \past{i} \delta (1-m)(1-\qut{i})(1-\kappa)\,, \nn
  \pastt{i} &=& \puit{i}[r_i(t) m \mu + (1-r_i(t))\mu]
                + \pait{i}[\delta m \mu+(1-\delta)\mu] \label{eq:MMCApas} \\
                && \mbox{} + \pust{i}[r_i(t) m \qat{i}+(1-r_i(t))\qat{i}]
                + \past{i}[\delta m \qat{i}+(1-\delta)\qat{i}]\,, \nn
  \paitt{i} &=& \puit{i}[r_i(t) m (1-\mu) + r_i(t)(1-m)(1-\mu)\kappa + (1-r_i(t))(1-\mu)] \label{eq:MMCApai} \\
                && \mbox{} + \pait{i} [\delta m (1-\mu) + \delta(1-m)(1-\mu)\kappa + (1-\delta)(1-\mu)] \nn
                && \mbox{} + \pust{i} [r_i(t) m (1-\qat{i}) + r_i(t)(1-m)(1-\qut{i})\kappa + (1-r_i(t))(1-\qat{i})] \nn
                && \mbox{} + \past{i} [\delta m (1-\qat{i}) + \delta(1-m)(1-\qut{i})\kappa + (1-\delta)(1-\qat{i})]\,. \nonumber
\eea
\end{widetext}
The probabilities of not being informed ($r_i(t)$) or infected ($\qut{i}$ and $\qat{i}$) by any neighbor, may be written as follows:
\bea
  r_i(t)  &=& \prod_j \left( 1-a_{ji}\pawt{j}\lambda  \right)\,,  \label{eq:ri} \\
  \qut{i} &=& \prod_j \left( 1-b_{ji}\pinft{j}\betau \right)\,,  \label{eq:qui} \\
  \qat{i} &=& \prod_j \left( 1-b_{ji}\pinft{j}\betaa \right)\,,  \label{eq:qai}
\eea
where $a_{ji}$ and $b_{ji}$ are the elements of the adjacency matrices of the UAU and SIS layer, respectively; and $\betaa = \gamma\betau$. In Eqs.~(\ref{eq:ri}) to~(\ref{eq:qai}), we assume independence of the probabilities of becoming infected or aware by any neighbor, which is the only hypothesis in the MMCA equations. Of course, the normalization condition
\beq
  \pust{i} + \puit{i} + \past{i} + \pait{i} = 1
\eeq
holds for all time steps.

Solving iteratively the system of Eqs.~(\ref{eq:MMCApus}) to~(\ref{eq:MMCApai}), together with Eqs.~(\ref{eq:ri}) to~(\ref{eq:qai}), we can track the time evolution of the awareness and the epidemics for any initial condition. Moreover, interestingly, we can solve analytically the stationary state of the full system, and determine the onset of the epidemics as a function of the rest of the parameters of the model.

\section{The onset of the epidemics in the presence of local and global awareness}

Starting out from the MMCA equations derived from the transition probability trees, one can calculate the critical epidemic threshold $\betau_c$ as a function of the rest of the parameters in the system at the stationary state $p_i(t+1)=p_i(t)$ for all nodes $i$ and states. First, since this epidemic threshold is given by the order parameter $\rhoinf$, which corresponds to the fraction of infected nodes in the system and is calculated as
\beq
  \rhoinf = \frac{1}{N}\sum_{i=1}^{N} \pinf{i} = \frac{1}{N}\sum_{i=1}^{N} (\pui{i}+\pai{i})\,,
\eeq
it is useful to add Eqs.~(\ref{eq:MMCApui}) and~(\ref{eq:MMCApai}) to obtain, in the steady state,
\bea
  &&\pinf{i} = \pinf{i} (1-\mu) \label{eq:MMCApinf} \\
                && \mbox{} + \pus{i} [r_i(1-m)(1-\qu{i}) + (1-r_i(1-m))(1-\qa{i})] \nn
                && \mbox{} + \pas{i} [\delta(1-m)(1-\qu{i}) + (1-\delta(1-m))(1-\qa{i})]\,. \nonumber
\eea

Near the onset of the epidemics, the probability of nodes to be infected is close to zero, i.e.\ $\pinf{i}= \epsilon_i \ll 1$. Accordingly,
Eqs.~(\ref{eq:qui}) and~(\ref{eq:qai}) are approximated as:
\bea
  \qu{i} &\approx& 1-\betau \sum_j b_{ji} \epsilon_j = 1 - \sigma_i\,,  \\
  \qa{i} &\approx& 1-\gamma\betau \sum_j b_{ji} \epsilon_j = 1 - \gamma\sigma_i\,,
\eea
where
\beq
  \sigma_i = \betau \sum_j b_{ji} \epsilon_j\,,
\eeq
and Eq.~(\ref{eq:MMCApinf}) becomes
\bea
  \epsilon_i &=& \epsilon_i (1-\mu) \label{eq:MMCApinfapprox} \\
                && \mbox{} + \pus{i} [r_i(1-m)\sigma_i + (1-r_i(1-m))\gamma\sigma_i] \nn
                && \mbox{} + \pas{i} [\delta(1-m)\sigma_i + (1-\delta(1-m))\gamma\sigma_i] \nn
             &=& \epsilon_i (1-\mu) \nn
                && \mbox{} + [\pu{i} r_i(1-m) + \pa{i}\delta(1-m)] \sigma_i \nn
                && \mbox{} + [\pu{i} (1-r_i(1-m)) + \pa{i}(1-\delta(1-m))] \gamma\sigma_i\,. \nonumber
\eea
Here we have made use of $\pu{i}=\pus{i}+\pui{i}\approx\pus{i}$ and $\pa{i}=\pas{i}+\pai{i}\approx\pas{i}$, since $\epsilon_i=\pui{i}+\pai{i}\ll 1$. In a similar way, removing $O(\epsilon_i)$ terms in the stationary state of Eqs.~(\ref{eq:MMCApus}) and~(\ref{eq:MMCApas}) we get
\bea
  \pu{i} &=& \pu{i} r_i(1-m) + \pa{i}\delta(1-m)\,, \label{eq:UAUpu} \\
  \pa{i} &=& \pu{i} (1-r_i(1-m)) + \pa{i}(1-\delta(1-m))\,. \label{eq:UAUpa}
\eea
These last equations correspond to an UAU process with mass media decoupled from the SIS process, with $\pu{i}+\pa{i}=1$. Substituting them in Eq.~(\ref{eq:MMCApinfapprox}) leads to
\bea
  \epsilon_i &=& (1-\mu)\epsilon_i + \pu{i}\sigma_i + \pa{i}\gamma\sigma_i \nn
             &=& (1-\mu)\epsilon_i + (\pu{i} + \pa{i}\gamma)\betau\sum_j b_{ji} \epsilon_j\,,
\eea
which can be written as
\beq
  \sum_j\left[ \betau(\pu{i}+\gamma\pa{i})b_{ji}-\mu\delta_{ij} \right] \epsilon_j\,,
  \label{eq:eigeneq}
\eeq
where $\delta_{ij}$ are the elements of the identity matrix. Defining matrix $H$ with elements
\beq
  h_{ij} = (\pu{i}+\gamma\pa{i})b_{ji}\,,
  \label{eq:hmatrix}
\eeq
the non-trivial solutions of Eq.~(\ref{eq:eigeneq}) are eigenvectors of $H$, whose eigenvalues are equal to $\mu / \betau$. Therefore, the onset of the epidemics is given by the largest eigenvalue of $H$,
\beq
  \betau_c = \frac{\mu}{\Lambda_{\mbox{\sz max}}(H)} \,.
\eeq
Note that matrix $H$ depends on the solutions of Eqs.~(\ref{eq:UAUpu}) and~(\ref{eq:UAUpa}), or equivalently
\beq
  \pa{i} = (1-\pa{i}) (1-r_i(1-m)) + \pa{i}(1-\delta(1-m))\,, \label{eq:UAU}
\eeq
where
\beq
  r_i = \prod_j \left( 1-a_{ji}\pa{j}\lambda \right), \label{eq:UAUri}
\eeq
which are also solved by iteration.

\section{Results}

\begin{figure}[tb]
\begin{center}
\includegraphics[width=\columnwidth,clip=]{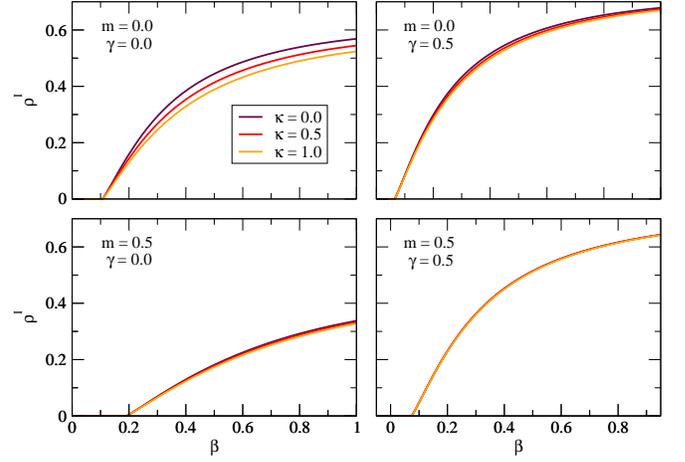}
\end{center}
\caption{(color online). Fraction of infected nodes as a function of the infectivity parameter $\beta$, for different values of the parameter $\kappa$. The networks used in this setup are the same as throughout the document. The rest of the values of parameters are: $\lambda=0.3$, $\delta=0.6$, $\mu=0.4$. The panel shows: top-left corner $\gamma=0.0$ i.e.\ total immunization and the mass media effect is turned off $m=0$; top-right corner the immunization is reduced $\gamma=0.5$; bottom-left corner mass media effect active $m=0.5$ with total immunization $\gamma=0.0$; and bottom-right corner mass media active and partial immunization $\gamma=0.5$.}
\label{fig:kappa_variable}
\end{figure}

\begin{figure}[tb]
\begin{center}
\includegraphics[width=\columnwidth,clip=]{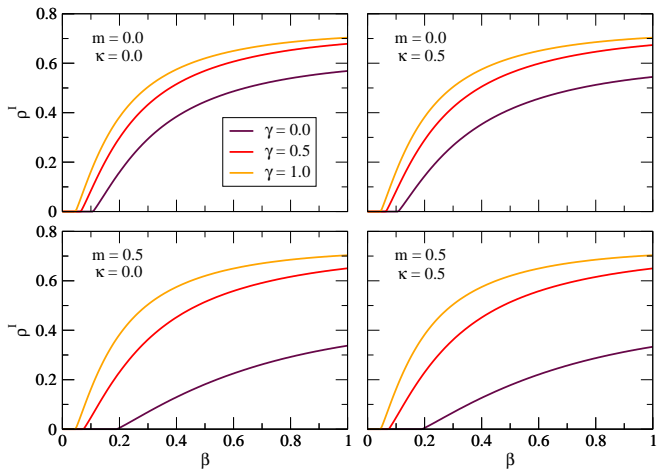}
\end{center}
\caption{(color online). Fraction of infected nodes as a function of the infectivity parameter $\beta$, for different values of the parameter $\gamma$. The networks used in this setup are the same as throughout the document. The rest of the values of parameters are set to $\lambda=0.3$, $\delta=0.6$ and $\mu=0.4$. The top-left panel shows the results for inexistent mass media and self-awareness; top-right has an intermediate self-awareness $\kappa=0.5$ keeping mass media turned off; bottom-left has no self-awareness but an intermediate mass media effect $m=0.5$ and bottom-right is set to intermediate values of both $m$ and $\kappa$.}
\label{fig:efecte_gamma}
\end{figure}

\begin{figure}[tb]
\begin{center}
\includegraphics[width=\columnwidth,clip=]{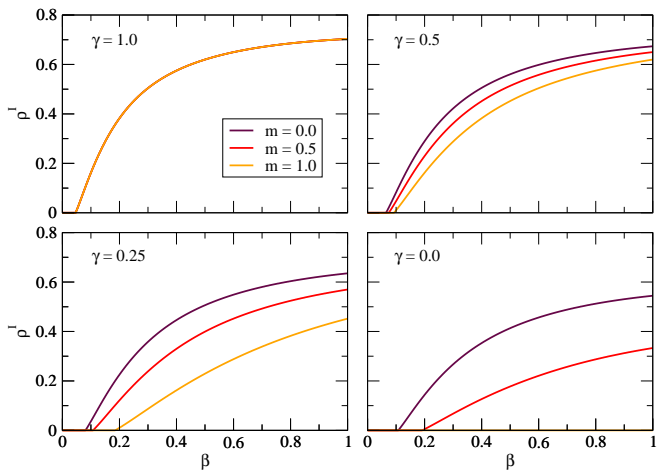}
\end{center}
\caption{(color online). Fraction of infected nodes as a function of the infectivity parameter $\beta$, for different values of the parameter representing the mass media effect, $m$. The networks used in this setup are the same as throughout the document. The rest of the values of parameters are: $\lambda=0.3$, $\delta=0.6$, $\mu=0.4$ and $\kappa$ is fixed to 0.5. The four panels correspond to values of $\gamma=1$, $\gamma=0.50$, $\gamma=0.25$ and $\gamma=0$.}
\label{fig:mass_variable}
\end{figure}

\begin{figure}[tb]
\begin{center}
\includegraphics[width=\columnwidth,clip=]{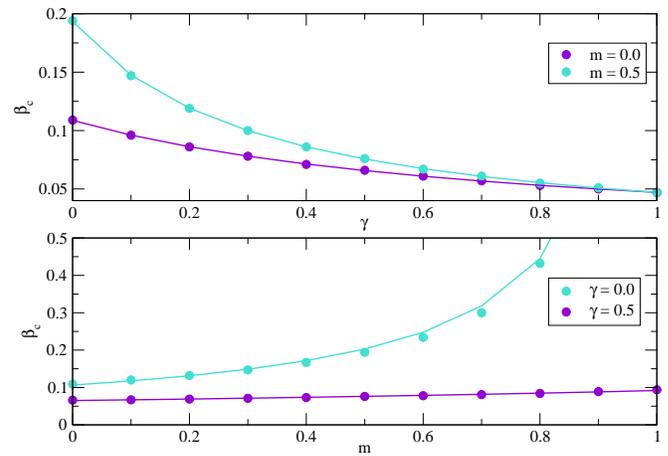}
\end{center}
\caption{(color online). Plot of $\beta_c$ as a function of $\gamma$ (top) and $m$ (bottom). Dots represent the data and the line is the fitting function \ $\beta_c \sim (a+bx)^{-1}$. As described, the less intense the immunization degree (larger $\gamma$) the lower the critical point of the epidemics, and conversely, the larger the intensity of the mass media $m$ the larger the critical onset.}
\label{fig:betafit}
\end{figure}


\
Here we investigate the effects of the three main parameters of the model: $\gamma$ (degree of immunization), $\kappa$ (self-awareness) and $m$ (mass media). The multiplex network we use is the following: the bottom layer corresponding to the physical contacts network is a power-law degree distribution network generated with the configurational model with an exponent of 2.5 and a size of 1000 nodes. The top layer representing the information contacts is the same network with 400 additional (non-overlapping with previous) links. This setup is the same as in \cite{granell13} for the sake of consistency. In the following, we analyze the incidence of the epidemics for different values of the parameters. To simplify the notation we will call $\beta$ to $\betau$.

In Fig.~\ref{fig:kappa_variable} we plot the fraction of infected nodes as a function of $\beta$ for different values of the self-awareness parameter $\kappa$. The rest of the parameters are set to intermediate values (see caption). The $\kappa$ parameter regulates the probability to be aware of your own disease, that is, the probability of going from UI state to AI state. In the figure panel we consider that the mass media is inactive (top) or active (bottom), and also that the immunization is total (left) or partial (right). Observing the figures, we see that by varying $\kappa$ the onset of the epidemics is not affected in any of the scenarios, which is a consequence of the absence of $\kappa$ in the Eqs.~(\ref{eq:hmatrix}) to~(\ref{eq:UAUri}) to determine the epidemic threshold.  We also observe no significant change on the final incidence. This result indicates that the incidence of the self-awareness in the whole process is negligible, even in the limiting cases where infected unaware individuals remain unaware of its sickness ($\kappa=0$) or certainly become aware of it ($\kappa=1$). The message extracted from these findings is that the self-awareness is not a key factor for the dynamical behavior of our system.

In Fig.~\ref{fig:efecte_gamma} we plot the density of infected nodes for different values of $\gamma$ for four combinations of the parameters $m$ and $\kappa$. The panel depicts the scenarios where the mass media is inactive (top) or active (bottom), and also that the self awareness is non-existent (left) or existent (right). The parameter $\gamma$ accounts for the immunity that a node gains when it is aware of the disease. We observe that for low values of $\gamma$ (high immunity) the final incidence of the epidemics is lowered, and the critical point is shifted right (comparing with non-existent coupling $\gamma=1$), for whatever the values of $m$ and $\kappa$. We can also see, if we compare the left and right plots, that $\kappa$ does not change the onset of the epidemics, and only slightly the final incidence, as explained previously.

To analyze the effect of the mass media on this model we also plot the density of infected nodes for different values of $m$, see Fig.~\ref{fig:mass_variable}. We fix the value of $\kappa$ and move only the immunization parameter $\gamma$. From this picture we observe that high values of $m$ shift the onset of the epidemics right and lower the final incidence. For low values of $\gamma$ the mass media effect is very pronounced (see bottom-right panel); on the other hand, when $\gamma=1$ (top-left plot) the mass media has no effect whatsoever as the epidemic layer has effectively been disconnected from the information layer.

We have shown how the critical point is affected by the parameters $\gamma$ and $m$. In Fig.~\ref{fig:betafit} we plot the non-linear dependence of $\beta_c$ on the degree of immunization and the mass media. Surprisingly enough, in both cases the dependence can be empirically fitted to an expression that is inversely linear to the parameters, i.e.\ $\beta_c \sim (a+bx)^{-1}$ for certain constants $a$ and $b$ and being the variable $x=\gamma$ or $x=m$.

\begin{figure}[tb]
\begin{center}
\includegraphics[width=\columnwidth,clip=]{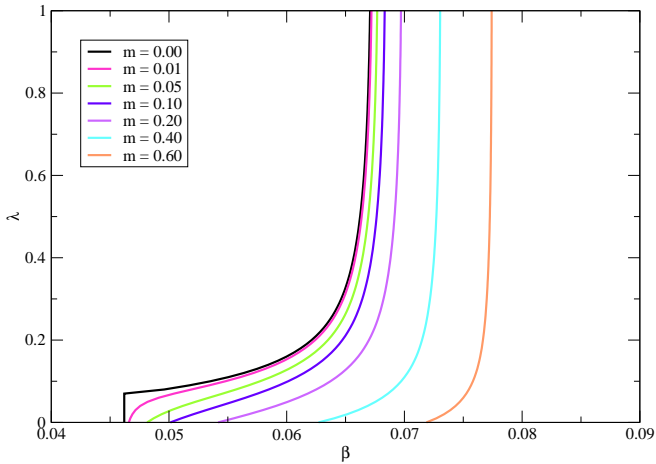}
\end{center}
\caption{(color online). Value of the onset of the epidemics as a function of the UAU parameter $\lambda$, for different values of the mass media parameter $m$. The setup is the same used throughout this document, with $\delta=0.6$, $\mu=0.4$ and parameters $\kappa=1.0$ and $\gamma=0.0$ which imply maximum coupling between layers.}
\label{fig:betac}
\end{figure}

The most remarkable outcome of the mass media effect is observed when representing the curve of critical points in the $\lambda-\beta$ phase space. As seen in the previous work, the coupling of the UAU and SIS layer implies the existence of a metacritical point, a point from which on the critical onset of the epidemics depends on the incidence of awareness in the population. The effect of the mass media is that of making the metacritical point vanish, this can be directly observed in Fig.\ref{fig:betac}. When $m=0$ we see that for low values of the rumor infectivity parameter $\lambda$, the onset of the epidemics $\beta_c$ is independent of $\lambda$, and it is not until a certain point (the metacritical point) that the UAU process begins to influence the onset of the epidemics. However, when the mass media is greater than zero (even for very small values) the phenomenology is different and the metacritical point disappears. The explanation of this phenomenon is rooted on the nature of the awareness provided by the mass media. Being it a random process acting on the whole population at each time step, whatever the capability of the awareness to spread it, a certain finite fraction of aware individuals will survive. This pool of aware individuals effectively decrease the epidemics, although non in a uniform linearly predictable way.


\section{Conclusions}

We have presented an extended analysis of a generalization of a model of competing spreading processes on multiplex networks. The original model \cite{granell13} accounts for the interplay between awareness and disease, both spreading processes competing on the same nodes, but on two different connectivity layers. The main physical result on such system relies on the emergence of a metacritical point, where the critical onsets of both dynamics get intertwined and the onset of the epidemics starts depending on the incidence of aware individuals. This result was obtained assuming that infected individuals get immediately aware, that aware individuals get immediately immunized, and that no awareness massive broadcast was present (mass media). Here we have relaxed the two first assumptions, and included the presence of a mass media aimed to check the validity of the previous results. The results are interesting, and while the immediacy of awareness when infected (self-awareness) has almost no effect on the dynamics, the other two factors, namely the degree of immunization of aware individuals and the mass media, do change the critical aspects of the epidemics spreading. Summarizing, we have found analytic expressions using a Microscopic Markov Chain Approach, that relate the decrease of the epidemic incidence with the increase in the level of immunization, and the modification of the incidence due to the mass media. The non-linear character of these relationships make the analytical approach extremely useful to understand the different scenarios.

\section{Acknowledgements}
This work has been partially supported by MINECO through Grant FIS2012-38266; and by the EC FET-Proactive Project PLEXMATH (grant 317614). A.A.\ also acknowledges partial financial support from the ICREA Academia and the James S.\ McDonnell Foundation.


\begin{thebibliography}{24}
\expandafter\ifx\csname natexlab\endcsname\relax\def\natexlab#1{#1}\fi
\expandafter\ifx\csname bibnamefont\endcsname\relax
  \def\bibnamefont#1{#1}\fi
\expandafter\ifx\csname bibfnamefont\endcsname\relax
  \def\bibfnamefont#1{#1}\fi
\expandafter\ifx\csname citenamefont\endcsname\relax
  \def\citenamefont#1{#1}\fi
\expandafter\ifx\csname url\endcsname\relax
  \def\url#1{\texttt{#1}}\fi
\expandafter\ifx\csname urlprefix\endcsname\relax\def\urlprefix{URL }\fi
\providecommand{\bibinfo}[2]{#2}
\providecommand{\eprint}[2][]{\url{#2}}

\bibitem[{\citenamefont{Newman}(2010)}]{Newman2010}
\bibinfo{author}{\bibfnamefont{M.~E.~J.} \bibnamefont{Newman}},
  \emph{\bibinfo{title}{Networks: An Introduction}} (\bibinfo{publisher}{Oxford
  University Press}, \bibinfo{year}{2010}).

\bibitem[{\citenamefont{Kivel\"{a} et~al.}(2013)\citenamefont{Kivel\"{a},
  Arenas, Barthelemy, Gleeson, Moreno, and Porter}}]{2013arXiv1309.7233K}
\bibinfo{author}{\bibfnamefont{M.}~\bibnamefont{Kivel\"{a}}},
  \bibinfo{author}{\bibfnamefont{A.}~\bibnamefont{Arenas}},
  \bibinfo{author}{\bibfnamefont{M.}~\bibnamefont{Barthelemy}},
  \bibinfo{author}{\bibfnamefont{J.~P.} \bibnamefont{Gleeson}},
  \bibinfo{author}{\bibfnamefont{Y.}~\bibnamefont{Moreno}}, \bibnamefont{and}
  \bibinfo{author}{\bibfnamefont{M.~A.} \bibnamefont{Porter}},
  \bibinfo{journal}{arXiv e-prints}  (\bibinfo{year}{2013}),
  \eprint{1309.7233}.

\bibitem[{\citenamefont{Kurant and Thiran}(2006)}]{kurant06}
\bibinfo{author}{\bibfnamefont{M.}~\bibnamefont{Kurant}} \bibnamefont{and}
  \bibinfo{author}{\bibfnamefont{P.}~\bibnamefont{Thiran}},
  \bibinfo{journal}{Phys. Rev. Lett.} \textbf{\bibinfo{volume}{96}},
  \bibinfo{pages}{138701} (\bibinfo{year}{2006}).

\bibitem[{\citenamefont{Mucha et~al.}(2010)\citenamefont{Mucha, Richardson,
  Macon, Porter, and Onnela}}]{mucha10}
\bibinfo{author}{\bibfnamefont{P.~J.} \bibnamefont{Mucha}},
  \bibinfo{author}{\bibfnamefont{T.}~\bibnamefont{Richardson}},
  \bibinfo{author}{\bibfnamefont{K.}~\bibnamefont{Macon}},
  \bibinfo{author}{\bibfnamefont{M.~A.} \bibnamefont{Porter}},
  \bibnamefont{and} \bibinfo{author}{\bibfnamefont{J.-P.}
  \bibnamefont{Onnela}}, \bibinfo{journal}{Science}
  \textbf{\bibinfo{volume}{328}}, \bibinfo{pages}{876} (\bibinfo{year}{2010}).

\bibitem[{\citenamefont{Szell et~al.}(2010)\citenamefont{Szell, Lambiotte, and
  Thurner}}]{Szell10}
\bibinfo{author}{\bibfnamefont{M.}~\bibnamefont{Szell}},
  \bibinfo{author}{\bibfnamefont{R.}~\bibnamefont{Lambiotte}},
  \bibnamefont{and} \bibinfo{author}{\bibfnamefont{S.}~\bibnamefont{Thurner}},
  \bibinfo{journal}{Proceedings of the National Academy of Sciences}
  \textbf{\bibinfo{volume}{107}}, \bibinfo{pages}{13636}
  (\bibinfo{year}{2010}).

\bibitem[{\citenamefont{G\'{o}mez-Garde\~{n}es
  et~al.}(2012)\citenamefont{G\'{o}mez-Garde\~{n}es, Reinares, Arenas, and
  Flor\'{\i}a}}]{gardenes12}
\bibinfo{author}{\bibfnamefont{J.}~\bibnamefont{G\'{o}mez-Garde\~{n}es}},
  \bibinfo{author}{\bibfnamefont{I.}~\bibnamefont{Reinares}},
  \bibinfo{author}{\bibfnamefont{A.}~\bibnamefont{Arenas}}, \bibnamefont{and}
  \bibinfo{author}{\bibfnamefont{L.~M.~M.} \bibnamefont{Flor\'{\i}a}},
  \bibinfo{journal}{Scientific reports} \textbf{\bibinfo{volume}{2}}
  (\bibinfo{year}{2012}).

\bibitem[{\citenamefont{Baxter et~al.}(2012)\citenamefont{Baxter, Dorogovtsev,
  Goltsev, and Mendes}}]{baxter12}
\bibinfo{author}{\bibfnamefont{G.~J.} \bibnamefont{Baxter}},
  \bibinfo{author}{\bibfnamefont{S.~N.} \bibnamefont{Dorogovtsev}},
  \bibinfo{author}{\bibfnamefont{A.~V.} \bibnamefont{Goltsev}},
  \bibnamefont{and} \bibinfo{author}{\bibfnamefont{J.~F.~F.}
  \bibnamefont{Mendes}}, \bibinfo{journal}{Phys. Rev. Lett.}
  \textbf{\bibinfo{volume}{109}}, \bibinfo{pages}{248701}
  (\bibinfo{year}{2012}).

\bibitem[{\citenamefont{Cozzo et~al.}(2012)\citenamefont{Cozzo, Arenas, and
  Moreno}}]{cozzo12}
\bibinfo{author}{\bibfnamefont{E.}~\bibnamefont{Cozzo}},
  \bibinfo{author}{\bibfnamefont{A.}~\bibnamefont{Arenas}}, \bibnamefont{and}
  \bibinfo{author}{\bibfnamefont{Y.}~\bibnamefont{Moreno}},
  \bibinfo{journal}{Phys. Rev. E} \textbf{\bibinfo{volume}{86}},
  \bibinfo{pages}{036115} (\bibinfo{year}{2012}).

\bibitem[{\citenamefont{Bianconi}(2013)}]{bianconi13}
\bibinfo{author}{\bibfnamefont{G.}~\bibnamefont{Bianconi}},
  \bibinfo{journal}{Phys. Rev. E} \textbf{\bibinfo{volume}{87}},
  \bibinfo{pages}{062806} (\bibinfo{year}{2013}).

\bibitem[{\citenamefont{G\'{o}mez et~al.}(2013)\citenamefont{G\'{o}mez,
  D\'{i}az-Guilera, G\'{o}mez-Garde\~nes, Vicente, Moreno, and
  Arenas}}]{gomez13}
\bibinfo{author}{\bibfnamefont{S.}~\bibnamefont{G\'{o}mez}},
  \bibinfo{author}{\bibfnamefont{A.}~\bibnamefont{D\'{i}az-Guilera}},
  \bibinfo{author}{\bibfnamefont{J.}~\bibnamefont{G\'{o}mez-Garde\~nes}},
  \bibinfo{author}{\bibfnamefont{C.~J..} \bibnamefont{Perez-Vicente}},
  \bibinfo{author}{\bibfnamefont{Y.}~\bibnamefont{Moreno}}, \bibnamefont{and}
  \bibinfo{author}{\bibfnamefont{A.}~\bibnamefont{Arenas}},
  \bibinfo{journal}{Physical Review Letters} \textbf{\bibinfo{volume}{110}},
  \bibinfo{pages}{028701} (\bibinfo{year}{2013}).

\bibitem[{\citenamefont{Sol\'e-Ribalta
  et~al.}(2013)\citenamefont{Sol\'e-Ribalta, De~Domenico, Kouvaris,
  D\'{i}az-Guilera, G\'omez, and Arenas}}]{soleribalta13}
\bibinfo{author}{\bibfnamefont{A.}~\bibnamefont{Sol\'e-Ribalta}},
  \bibinfo{author}{\bibfnamefont{M.}~\bibnamefont{De~Domenico}},
  \bibinfo{author}{\bibfnamefont{N.~E.} \bibnamefont{Kouvaris}},
  \bibinfo{author}{\bibfnamefont{A.}~\bibnamefont{D\'{i}az-Guilera}},
  \bibinfo{author}{\bibfnamefont{S.}~\bibnamefont{G\'omez}}, \bibnamefont{and}
  \bibinfo{author}{\bibfnamefont{A.}~\bibnamefont{Arenas}},
  \bibinfo{journal}{Phys. Rev. E} \textbf{\bibinfo{volume}{88}},
  \bibinfo{pages}{032807} (\bibinfo{year}{2013}).

\bibitem[{\citenamefont{Radicchi and Arenas}(2013)}]{radicchi13}
\bibinfo{author}{\bibfnamefont{F.}~\bibnamefont{Radicchi}} \bibnamefont{and}
  \bibinfo{author}{\bibfnamefont{A.}~\bibnamefont{Arenas}},
  \bibinfo{journal}{Nature Physics} \textbf{\bibinfo{volume}{9}},
  \bibinfo{pages}{717} (\bibinfo{year}{2013}).

\bibitem[{\citenamefont{De~Domenico et~al.}(2013)\citenamefont{De~Domenico,
  Sol\'e-Ribalta, Cozzo, Kivel\"a, Moreno, Porter, G\'omez, and
  Arenas}}]{dedomenico13}
\bibinfo{author}{\bibfnamefont{M.}~\bibnamefont{De~Domenico}},
  \bibinfo{author}{\bibfnamefont{A.}~\bibnamefont{Sol\'e-Ribalta}},
  \bibinfo{author}{\bibfnamefont{E.}~\bibnamefont{Cozzo}},
  \bibinfo{author}{\bibfnamefont{M.}~\bibnamefont{Kivel\"a}},
  \bibinfo{author}{\bibfnamefont{Y.}~\bibnamefont{Moreno}},
  \bibinfo{author}{\bibfnamefont{M.~A.} \bibnamefont{Porter}},
  \bibinfo{author}{\bibfnamefont{S.}~\bibnamefont{G\'omez}}, \bibnamefont{and}
  \bibinfo{author}{\bibfnamefont{A.}~\bibnamefont{Arenas}},
  \bibinfo{journal}{Phys. Rev. X} \textbf{\bibinfo{volume}{3}},
  \bibinfo{pages}{041022} (\bibinfo{year}{2013}).

\bibitem[{\citenamefont{Sahneh et~al.}(2013)\citenamefont{Sahneh, Scoglio, and
  Mieghem}}]{SahnehSM13}
\bibinfo{author}{\bibfnamefont{F.~D.} \bibnamefont{Sahneh}},
  \bibinfo{author}{\bibfnamefont{C.~M.} \bibnamefont{Scoglio}},
  \bibnamefont{and} \bibinfo{author}{\bibfnamefont{P.~V.}
  \bibnamefont{Mieghem}}, \bibinfo{journal}{IEEE/ACM Trans. Netw.}
  \textbf{\bibinfo{volume}{21}}, \bibinfo{pages}{1609} (\bibinfo{year}{2013}).

\bibitem[{\citenamefont{{Sanz} et~al.}(2014)\citenamefont{{Sanz}, {Xia},
  {Meloni}, and {Moreno}}}]{sanz2014}
\bibinfo{author}{\bibfnamefont{J.}~\bibnamefont{{Sanz}}},
  \bibinfo{author}{\bibfnamefont{C.-Y.} \bibnamefont{{Xia}}},
  \bibinfo{author}{\bibfnamefont{S.}~\bibnamefont{{Meloni}}}, \bibnamefont{and}
  \bibinfo{author}{\bibfnamefont{Y.}~\bibnamefont{{Moreno}}},
  \bibinfo{journal}{arXiv e-prints}  (\bibinfo{year}{2014}),
  \eprint{1402.4523}.

\bibitem[{\citenamefont{Buldyrev et~al.}(2010)\citenamefont{Buldyrev, Parshani,
  Paul, Stanley, and Havlin}}]{buldyrev10}
\bibinfo{author}{\bibfnamefont{S.~V.} \bibnamefont{Buldyrev}},
  \bibinfo{author}{\bibfnamefont{R.}~\bibnamefont{Parshani}},
  \bibinfo{author}{\bibfnamefont{G.}~\bibnamefont{Paul}},
  \bibinfo{author}{\bibfnamefont{H.~E.} \bibnamefont{Stanley}},
  \bibnamefont{and} \bibinfo{author}{\bibfnamefont{S.}~\bibnamefont{Havlin}},
  \bibinfo{journal}{Nature} \textbf{\bibinfo{volume}{464}},
  \bibinfo{pages}{1025} (\bibinfo{year}{2010}).

\bibitem[{\citenamefont{Gao et~al.}(2011)\citenamefont{Gao, Buldyrev, Stanley,
  and Havlin}}]{gao11}
\bibinfo{author}{\bibfnamefont{J.}~\bibnamefont{Gao}},
  \bibinfo{author}{\bibfnamefont{S.~V.} \bibnamefont{Buldyrev}},
  \bibinfo{author}{\bibfnamefont{H.~E.} \bibnamefont{Stanley}},
  \bibnamefont{and} \bibinfo{author}{\bibfnamefont{S.}~\bibnamefont{Havlin}},
  \bibinfo{journal}{Nature Physics} \textbf{\bibinfo{volume}{8}},
  \bibinfo{pages}{40} (\bibinfo{year}{2011}).

\bibitem[{\citenamefont{Funk et~al.}(2009)\citenamefont{Funk, Gilad, Watkins,
  and Jansen}}]{funk09}
\bibinfo{author}{\bibfnamefont{S.}~\bibnamefont{Funk}},
  \bibinfo{author}{\bibfnamefont{E.}~\bibnamefont{Gilad}},
  \bibinfo{author}{\bibfnamefont{C.}~\bibnamefont{Watkins}}, \bibnamefont{and}
  \bibinfo{author}{\bibfnamefont{V.~A.~A.} \bibnamefont{Jansen}},
  \bibinfo{journal}{Proceedings of the National Academy of Sciences}
  \textbf{\bibinfo{volume}{106}}, \bibinfo{pages}{6872} (\bibinfo{year}{2009}).

\bibitem[{\citenamefont{Granell et~al.}(2013)\citenamefont{Granell, G\'omez,
  and Arenas}}]{granell13}
\bibinfo{author}{\bibfnamefont{C.}~\bibnamefont{Granell}},
  \bibinfo{author}{\bibfnamefont{S.}~\bibnamefont{G\'omez}}, \bibnamefont{and}
  \bibinfo{author}{\bibfnamefont{A.}~\bibnamefont{Arenas}},
  \bibinfo{journal}{Phys. Rev. Lett.} \textbf{\bibinfo{volume}{111}},
  \bibinfo{pages}{128701} (\bibinfo{year}{2013}).

\bibitem[{\citenamefont{Meloni et~al.}(2011)\citenamefont{Meloni, Perra,
  Arenas, G\'{o}mez, Moreno, and Vespignani}}]{Meloni11}
\bibinfo{author}{\bibfnamefont{S.}~\bibnamefont{Meloni}},
  \bibinfo{author}{\bibfnamefont{N.}~\bibnamefont{Perra}},
  \bibinfo{author}{\bibfnamefont{A.}~\bibnamefont{Arenas}},
  \bibinfo{author}{\bibfnamefont{S.}~\bibnamefont{G\'{o}mez}},
  \bibinfo{author}{\bibfnamefont{Y.}~\bibnamefont{Moreno}}, \bibnamefont{and}
  \bibinfo{author}{\bibfnamefont{A.}~\bibnamefont{Vespignani}},
  \bibinfo{journal}{Scientific Reports} \textbf{\bibinfo{volume}{1}},
  \bibinfo{pages}{62} (\bibinfo{year}{2011}).

\bibitem[{\citenamefont{Paolotti et~al.}(2014)\citenamefont{Paolotti, Carnahan,
  Colizza, Eames, Edmunds, Gomes, Koppeschaar, Rehn, Smallenburg, Turbelin
  et~al.}}]{influenzanet}
\bibinfo{author}{\bibfnamefont{D.}~\bibnamefont{Paolotti}},
  \bibinfo{author}{\bibfnamefont{A.}~\bibnamefont{Carnahan}},
  \bibinfo{author}{\bibfnamefont{V.}~\bibnamefont{Colizza}},
  \bibinfo{author}{\bibfnamefont{K.}~\bibnamefont{Eames}},
  \bibinfo{author}{\bibfnamefont{J.}~\bibnamefont{Edmunds}},
  \bibinfo{author}{\bibfnamefont{G.}~\bibnamefont{Gomes}},
  \bibinfo{author}{\bibfnamefont{C.}~\bibnamefont{Koppeschaar}},
  \bibinfo{author}{\bibfnamefont{M.}~\bibnamefont{Rehn}},
  \bibinfo{author}{\bibfnamefont{R.}~\bibnamefont{Smallenburg}},
  \bibinfo{author}{\bibfnamefont{C.}~\bibnamefont{Turbelin}},
  \bibnamefont{et~al.}, \bibinfo{journal}{Clin. Microbiol. Infect.}
  \textbf{\bibinfo{volume}{20}}, \bibinfo{pages}{17} (\bibinfo{year}{2014}).

\bibitem[{\citenamefont{Colizza et~al.}(2007)\citenamefont{Colizza,
  Pastor-Satorras, and Vespignani}}]{Colizza:2007}
\bibinfo{author}{\bibfnamefont{V.}~\bibnamefont{Colizza}},
  \bibinfo{author}{\bibfnamefont{R.}~\bibnamefont{Pastor-Satorras}},
  \bibnamefont{and}
  \bibinfo{author}{\bibfnamefont{A.}~\bibnamefont{Vespignani}},
  \bibinfo{journal}{Nat Phys} \textbf{\bibinfo{volume}{3}},
  \bibinfo{pages}{276} (\bibinfo{year}{2007}).

\bibitem[{\citenamefont{Gleeson}(2011)}]{gleeson11}
\bibinfo{author}{\bibfnamefont{J.~P.} \bibnamefont{Gleeson}},
  \bibinfo{journal}{Phys. Rev. Lett.} \textbf{\bibinfo{volume}{107}},
  \bibinfo{pages}{068701} (\bibinfo{year}{2011}).

\bibitem[{\citenamefont{G\'{o}mez et~al.}(2010)\citenamefont{G\'{o}mez, Arenas,
  Borge-Holthoefer, Meloni, and Moreno}}]{gomez10}
\bibinfo{author}{\bibfnamefont{S.}~\bibnamefont{G\'{o}mez}},
  \bibinfo{author}{\bibfnamefont{A.}~\bibnamefont{Arenas}},
  \bibinfo{author}{\bibfnamefont{J.}~\bibnamefont{Borge-Holthoefer}},
  \bibinfo{author}{\bibfnamefont{S.}~\bibnamefont{Meloni}}, \bibnamefont{and}
  \bibinfo{author}{\bibfnamefont{Y.}~\bibnamefont{Moreno}},
  \bibinfo{journal}{Europhys. Lett.} \textbf{\bibinfo{volume}{89}},
  \bibinfo{pages}{38009} (\bibinfo{year}{2010}).

\end{thebibliography}

\end{document}